\documentclass[12pt]{article}
\usepackage{epsfig}
\newcommand\pt{p_{\rm \scriptscriptstyle T}}
\newcommand\mt{m_{\rm \scriptscriptstyle T}}

\begin{document}
\begin{titlepage}
\nopagebreak
{\flushright{
        \begin{minipage}{4cm}
         UPRF-2003-10\hfill \\
         Bicocca-FT-03-14\hfill \\
%        IFUM/613-FT \hfill \\
%        LNF-98/008(P) \hfill \\
%        hep-ph/9803400\hfill \\
        \end{minipage}        }

}
\vfill
\begin{center}
{\LARGE 
{ \bf \sc
Charm Cross Sections for the\\[5pt] Tevatron Run II}}
\vskip .5cm
{\bf Matteo Cacciari
}
\\
\vskip .1cm
{Dipartimento di Fisica,
Universit\`a di Parma, Italy, and\\
INFN, Sezione di Milano, Gruppo Collegato di Parma}
\vskip .5cm
{\bf Paolo Nason}
\\
\vskip 0.1cm
{INFN, Sezione di Milano}
%\vskip .4cm
\end{center}
\nopagebreak
\vfill
%\vskip 3cm
\begin{abstract}
We present a calculation of the $D^{*+}$, $D^+$ and $D^0$ meson single inclusive
production cross section for the Tevatron Run II.
We use the FONLL approach in perturbative QCD, which, besides including
the known next-to-leading order corrections, also provides for the resummation 
at the next-to-leading logarithmic level
of terms enhanced at large $\pt$ by powers of $\log(\pt/m)$, where $m$ is the
charm mass and $\pt$ is its transverse momentum.
Non-perturbative effects in charm hadronization
are extracted, in moment space, from recent ALEPH data for $D^*$ 
fragmentation in $e^+e^-$ collisions.
\end{abstract}
\vskip 1cm
%CERN-TH/98-77 \hfill \\
%March 20, 1998 \hfill
\vfill
\end{titlepage}
\section{Introduction}
Heavy flavour production at hadron colliders has been mostly studied
for bottomed hadrons so far, due to the possibility of using
muons or $J/\Psi$ to tag the $B$ hadrons.
At CDF, thanks to the development of the silicon vertex trigger,
charm production studies have become possible. This opens
up the interesting possibilities of comparing QCD predictions
for charm production with experimental results, much in the same
way as one does for bottomed hadrons. One is then especially interested
in seeing whether the predicted charm cross section bears the
same relationship with respect to data as the bottom cross section.

In the present work, we provide cross sections for charm production
for the Tevatron Run II. The calculation of the cross section is performed
according to the FONLL prescription \cite{Cacciari:1998it},
and the effect of a non-perturbative
fragmentation function is included in full analogy with the work
of ref.~\cite{Cacciari:2002pa},
where a prediction for $B$ production at the Tevatron Run I
is given. We will thus not review the FONLL framework, and the moment
space method for the extraction of the non-perturbative fragmentation
function, rather referring the reader to refs.
\cite{Cacciari:1998it,Cacciari:2002pa}.\footnote{Suffice here to say that the perturbative FONLL calculation includes the full
next-to-leading order calculation for heavy quark production (and therefore also
the power suppressed $m/p_T$ terms, all important in the $p_T\sim m$ region), 
plus the resummation to next-to-leading logarithmic accuracy of the $\log(p_T/m)$ 
terms, which are large where $p_T \gg m$. The moment space method, on the other
hand, takes care of properly extracting from experimental data, and importing in
the process at hand, the relevant non-perturbative information.}
The only novel aspect of the present work that needs to be discussed
is the phenomenological extraction of the non-perturbative
$D$-meson fragmentation function, which will serve as input to
our calculation of the $D$ spectrum.
Good data on $D^*$ fragmentation have been published by the ALEPH
collaboration \cite{Barate:1999bg}, together with a study yielding
the relative fraction of directly produced $D$ and $D^*$
mesons. In section \ref{sec:dfromdstar}, we will first describe how
we have constructed the $D^0$ and $D^+$ fragmentation functions in terms
of the $D^*$ one and of the relative fractions
of directly produced vector and pseudoscalar states.
In section \ref{sec:dstarfromdata} we discuss how the $D^*$
non-perturbative fragmentation function itself was extracted
from the ALEPH data.
In section \ref{sec:pheno} we present our phenomenological predictions,
and in section \ref{sec:conc} we give our conclusions.

\section{Construction of the $D^0$ and $D^+$  non-perturbative fragmentation 
functions}%
\label{sec:dfromdstar}

Non-perturbative information concerning the hadronization of the charm quarks
into the observed $D$ mesons must be extracted from experimental data.
High quality data at present only exist 
for fragmentation into the vector state $D^*$, as measured by the 
ALEPH collaboration \cite{Barate:1999bg} at LEP. The extraction of this
non-perturbative fragmentation function (FF from now on)
will be described in the next section.
In the present one we show how this FF can be suitably modified in order 
to describe the production of the pseudoscalar states $D^0$ and $D^+$ too.

The $D^0$ FF has been postulated to be a linear
combination of $D^0$'s coming from the decays
of the $D^{*0}$'s and $D^{*+}$'s, and $D^0$'s not coming from $D^*$
decays, that we call here ``primary''$D^0$'s :
\begin{eqnarray} \nonumber
F(c\to D^0)&=& F_p(c\to D^0)
  + F(c\to D^{*+}) \otimes
 F(D^{*+}\to D^0)
\nonumber \\
& +& F(c\to D^{*0}) \otimes
 F(D^{*+}\to D^0)
\nonumber
\end{eqnarray}
where $F_p(c\to D^0)$ is the fragmentation function for primary
$D^0$ production, $F(c\to D^{*0})$ and $F(c\to D^{*+})$ are the 
fragmentation functions
for $D^*$ states production, $F(D^{*0}\to D^0)$ and $F(D^{*+}\to D^0)$ describe 
the decay of a $D^{*}$ into a $D^0$.

We assume that the production of primary $D$ and $D^*$ is flavour
independent, and the only flavour dependencies arise in the
decay of the $D^*$. In other words, one assumes as usual that isospin
is a good symmetry, except that, due to the known fact that the $D\pi$ decay
mode is barely accessible to the $D^*$, small isospin violations
are amplified in the decay.
Under this assumption, we can extract from the ALEPH paper~\cite{Barate:1999bg}
the primary $D^0$ fraction and the $D^{*+}$ and $D^{*0}$
fraction\footnote{Here and in the following, the branching fractions
are presented and used without an associated experimental uncertainty.
While such errors are in principle available from the experimental
papers, we have refrained from quoting them as their contribution to the
total uncertainty will turn out to be negligible.}:
\begin{equation}
{\rm BR}(c\to D^0_p) = 0.168,\quad{\rm BR}(c\to D^*)=0.233\; .
\end{equation}
Observe that these values conflict with the naive assumption
of spin counting, that would predict a ratio of the $D$ to $D^*$ states
equal to the ratio of the number of spin states, $1/3$.

We now need a model for the $D^*$ FF. For sake of definiteness,
we have taken the analytic forms of ref.~\cite{Braaten:1995bz}
for the fragmentation into pseudoscalar (P) and vector (V) states.
Such FF's will be denoted by
$D_{BCFY}^{(P)}(z;r)$ and $D_{BCFY}^{(V)}(z;r)$ respectively, and understood to
be normalized to one. A single
non-perturbative parameter $r$ describes both pseudoscalar and vector
production  in this formulation, and we can thus use the parameter fitted to
$D^*$ production as input in the pseudoscalar ($D^0$ and $D^+$) fragmentation
functions. We will comment in the following about the influence of this choice
on our result.

The $D^*$ 
decay into a $D^0$ could be calculated exactly using the known decay chains.
However, since the decay $D^*\to \pi D^0$ is dominant and
the decay products are nearly at rest in the decay frame,
the effect of this decay is simply modeled as a rescaling
of the longitudinal momentum by a factor $m_D/m_{D^*}$ in the FF for $D^*$
production:
\begin{eqnarray}
{\tilde D}_{BCFY}^{(V)}(z;r)&=&
%\int D_{BCFY}^{(V)}(z_{D^*};r) \;
%\delta\left(z - z_{D^*}\frac{m_D}{m_{D^*}}\right) dz_{D^*} \nonumber \\
\int dy\,dz_{D^*}\,\delta(z-yz_{D^*})D_{BCFY}^{(V)}(z_{D^*};r) \;
\delta\left(y-\frac{m_D}{m_{D^*}}\right)\nonumber \\
&=&\Theta\left(\frac{m_D}{m_{D^*}}-z\right)
D_{BCFY}^{(V)}\left(\frac{m_{D^*}}{m_D}z;r\right) \frac{m_{D^*}}{m_D} \; ,
\end{eqnarray}
$\Theta(x)$ being the Heaviside step function.
We can therefore write
\begin{eqnarray}
D^{c\to D^0}(z;r) &=&  {\rm BR}(c\to D^0_p) D_{BCFY}^{(P)}(z;r) + \nonumber \\
&+& {\rm BR}(c\to D^*)\big({\rm BR}(D^{*0}\to D^0)+
{\rm BR}(D^{*+}\to D^0)\big) {\tilde D}_{BCFY}^{(V)}(z;r)
\label{eq:d0ff}
\end{eqnarray}
where~\cite{Hagiwara:2002fs}
\begin{equation}
{\rm BR}(D^{*0}\to D^0) = 1,\quad{\rm BR}(D^{*+}\to D^0)=0.677 \; .
\end{equation}
Putting in numbers, eq.~(\ref{eq:d0ff}) reads
\begin{equation}
D^{c\to D^0}(z;r) =  0.168 D_{BCFY}^{(P)}(z;r) +
0.39 {\tilde D}_{BCFY}^{(V)}(z;r)
\label{eq:ffdzero}
\end{equation}
showing that about
$2/3$ of the fragmentation function depends in a precise way on
the $D^*$ fragmentation.

The FF for $D^+$ production can be constructed along similar lines. Since
$D^{*0}$ cannot decay to $D^{+}$ we only have
\begin{equation} 
F(c\to D^+)= F_p(c\to D^+)
  + F(c\to D^{*+}) \otimes
 F(D^{*+}\to D^+)
\end{equation}
and therefore
\begin{eqnarray}
D^{c\to D^+}(z;r) &=&  {\rm BR}(c\to D^+_p) D_{BCFY}^{(P)}(z;r) + \nonumber \\
&+& {\rm BR}(c\to D^*)
{\rm BR}(D^{*+}\to D^+){\tilde D}_{BCFY}^{(V)}(z;r)
\end{eqnarray}
with\footnote{Note that ${\rm BR}(c\to D^+_p) = 0.162$ is actually to be
considered equal to  ${\rm BR}(c\to D^0_p) = 0.168$ within the experimental
uncertainties and consistently with our flavour-independence
hypothesis.}~\cite{Barate:1999bg,Hagiwara:2002fs}
\begin{equation}
{\rm BR}(c\to D^+_p) = 0.162,\qquad {\rm BR}(D^{*+}\to D^+) = 0.323 \; .
\end{equation}
Numerically
\begin{equation}
D^{c\to D^+}(z;r) = 0.162  D_{BCFY}^{(P)}(z;r) 
+ 0.07153 {\tilde D}_{BCFY}^{(V)}(z;r)\;.
\label{eq:ffdplus}
\end{equation}
For the sake of completeness, we now write down the FF for production of $D^*$
states:
\begin{equation}
D^{c\to D^*}(z;r) = {\rm BR}(c\to D^*)D_{BCFY}^{(V)}(z;r)
=0.233 D_{BCFY}^{(V)}(z;r)\;.
\label{eq:ffdstar}
\end{equation}
The non-perturbative parameter $r$, which in the model of
ref.~\cite{Braaten:1995bz} can be
interpreted as the ratio of the constituent mass of the light quark to the mass
of the meson, will be more precisely determined in the next section by a
comparison with the ALEPH $D^*$ data.

It is worth noting now that, in the present work, the model
of ref.~\cite{Braaten:1995bz} is needed in order to compensate for the
lack of an explicit measurement of the direct component of the
$D^{+/0}$ fragmentation function. In the case of the $D^0$ production
the influence of the model is marginal, since most $D^0$ are produced
through $D^*$ decays. In the case of $D^+$ production we have the opposite
situation. Therefore, our $D^+$ spectrum will strongly reflect this model
dependence. New data on $D^+$ fragmentation may eventually help
in reducing this uncertainty in the future.

\section{Extraction of the $D^*$ non-perturbative fragmentation function}%
\label{sec:dstarfromdata}

\begin{figure}[t]
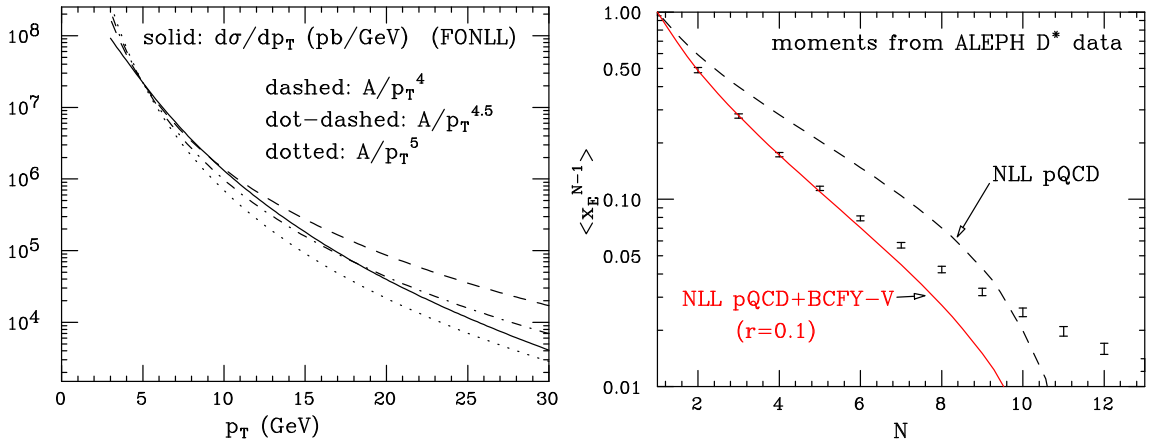

\begin{center}
\epsfig{file=ptdep.eps,height=0.37\textwidth}
\epsfig{file=aleph-mom.eps,height=0.37\textwidth}
\end{center}
\caption{\label{fig:ptdep}\small 
 a) Left plot: $p_T$ spectrum of the charm cross
section in $p\bar p$ collisions at $\sqrt{s} = 1960$ GeV, compared to $A/p_T^n$
behaviours. b) Right plot: Mellin moments of $D^*$ fragmentation distribution 
as calculated
from ALEPH data~\protect\cite{Barate:1999bg}, compared to pure perturbative QCD
and to pQCD convoluted with the non-perturbative form
$D_{BCFY,N}^{(V)}(r=0.1)$.}
\end{figure}

The non-perturbative fragmentation function has been extracted from $D^*$
fragmentation data from the ALEPH collaboration~\cite{Barate:1999bg}. 
We assume the measured energy fraction ($x_E$) distribution (normalized to one)
to be described by a convolution (a product, 
in Mellin moments space) of a
perturbative contribution (which describes the production of a $c$ quark in the
hard interaction) and the BCFY non-perturbative form describing its
hadronization into the $D^*$:
\begin{equation}
\sigma_N(c\to D^*)(r) = D_N^{pQCD} D_{BCFY,N}^{(V)}(r) \; .
\end{equation}
The non-perturbative parameter $r$ is then determined by 
comparing $\sigma_N$ to the measured $\langle x_E^{N-1}\rangle$ values.
 
According to the discussion  published in~\cite{Cacciari:2002pa}
about $B$ meson production at CDF, we have taken care that the fitted
non-perturbative fragmentation function properly describes the Mellin
moments of the $D^*$ spectrum in $e^+e^-$ experimental data around
$N=4,5$ rather than the whole $x_E$ shape, as this information is the most
relevant one when calculating the hadronic production.
Indeed, as shown in figure~\ref{fig:ptdep}(a), in the region of interest between
$p_T = 5$ and $p_T = 20$~GeV, the charm transverse momentum
distribution predicted by the FONLL calculation falls off like $A/p_T^n$, 
with $n$ between 4 and 5. 

Figure~\ref{fig:ptdep}(b) shows a comparison of the Mellin moments calculated
from the $D^*$ fragmentation data of the ALEPH 
Collaboration~\cite{Barate:1999bg} with the result of a convolution of a
perturbative and a non-perturbative distribution. The perturbative component
$D_N^{pQCD}$ 
resums to all orders and to next-to-leading logarithmic~\cite{Mele:1991cw} 
accuracy the large $\alpha_s^n \log^k(Q^2/m^2)~~(k\le n)$ terms ($Q$ being the
$e^+e^-$ centre-of-mass energy, and $m$ the heavy quark mass) which appear in
the fixed order calculation. This resummation is analogous to the one performed
in the FONLL result, where $\alpha_s^n \log^k(p_T^2/m^2)$ terms are resummed.
 For the non-perturbative distribution, the
functional form for vector states given by Braaten et al.~\cite{Braaten:1995bz}
has been chosen. Though no rigorous fit\footnote{The choice of not performing a
rigorous fit of the moments is related to the fact that their errors have been
only naively estimated from the errors of each $x_E$ bin, a full correlation
matrix being unavailable. Lacking a rigorous calculation for the errors on the
moments and  the correlations between the various moments, a fit would yield
results of little significance as far as the errors on the fitted parameters
are concerned.} of its free parameter $r$ has been performed, one can see from 
figure~\ref{fig:ptdep}(b) that the choice $r=0.1$ gives a good description of
the relevant moments around $N=4,5$. The failure of the curve to describe
higher-$N$ moments is of course related to unaccounted-for sub-leading 
logarithms
related to soft-gluon emission becoming more important. 
Resummation of these terms would improve the picture but, though
available~\cite{Cacciari:2001cw,Cacciari:2002xb}, it has not been included 
since it would not
affect this phenomenological application\footnote{By tweaking the factorization
scales which appear in the resummed perturbative calculation one could improve
the behaviour of the curve in the $N > 6$ region even without resumming the
soft-gluon logarithms. This implies a different
value for the non-perturbative parameter $r$ when fitting the experimental
moments. We have checked that the differences compensate, and the predictions
for $D$ meson production in $p\bar p$ collisions remain identical within a few
percent.}.

\section{Phenomenological results}
\label{sec:pheno}

\begin{figure}[t]
\epsfig{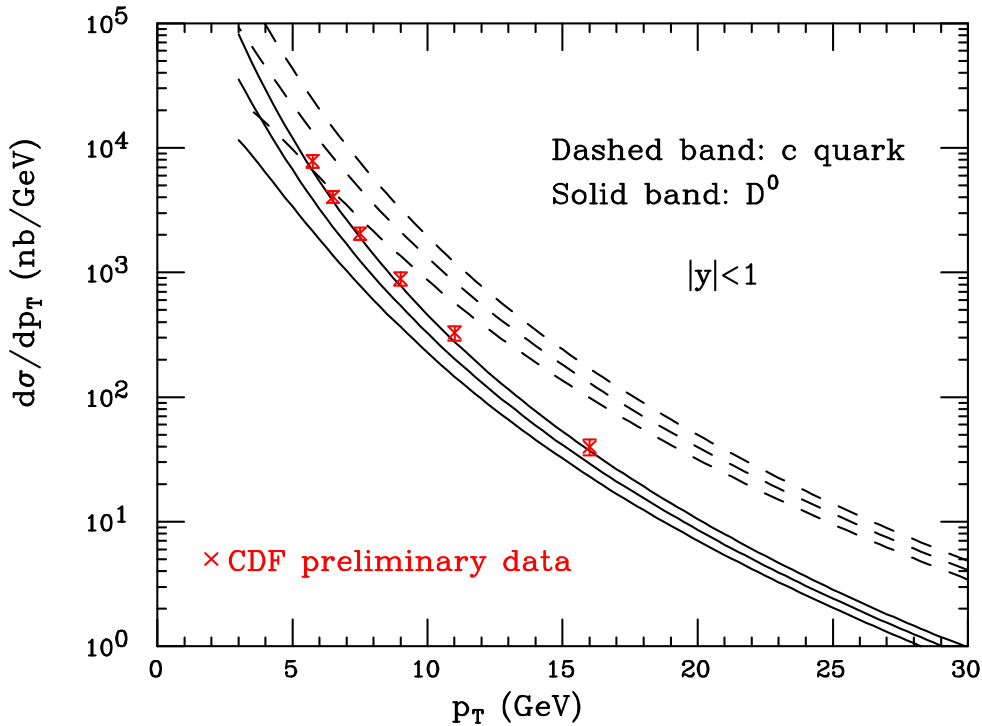}
\begin{center}
\caption{
QCD predictions for the $D^{0}$ differential cross
section at the Tevatron Run II. CDF preliminary data are also shown for
comparison.
}
\label{fig:d0run2}
\end{center}
\end{figure}
\begin{figure}[t]
\epsfig{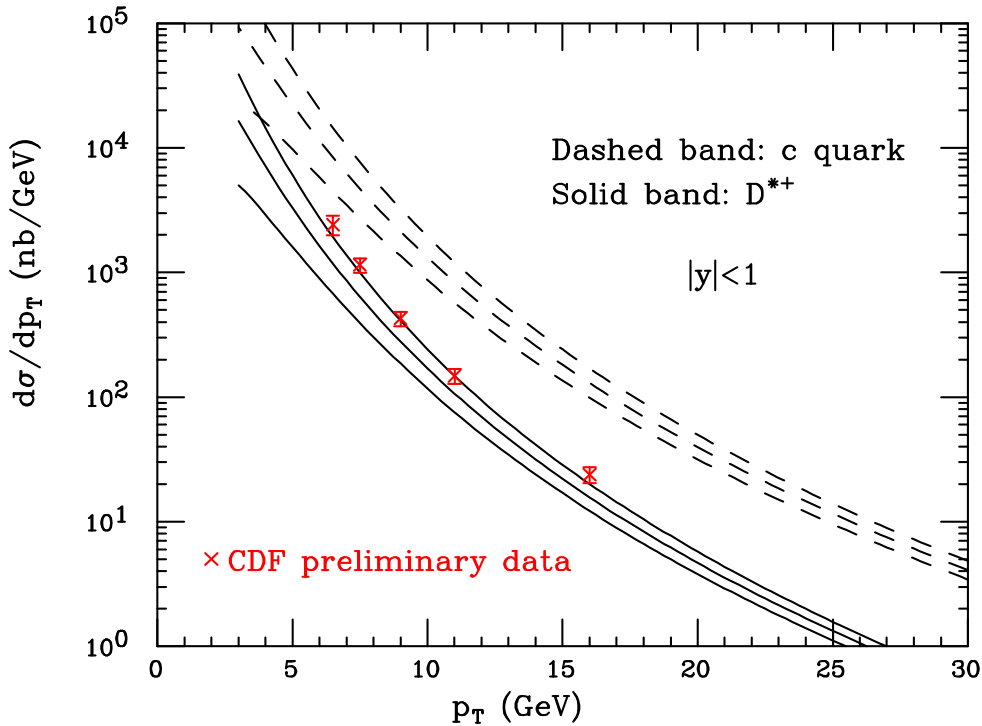}
\begin{center}
\caption{
Same as fig.~\protect\ref{fig:d0run2}, but for $D^{*+}$ production.
}
\label{fig:dstarplusrun2}
\end{center}
\end{figure}
\begin{figure}[t]
\epsfig{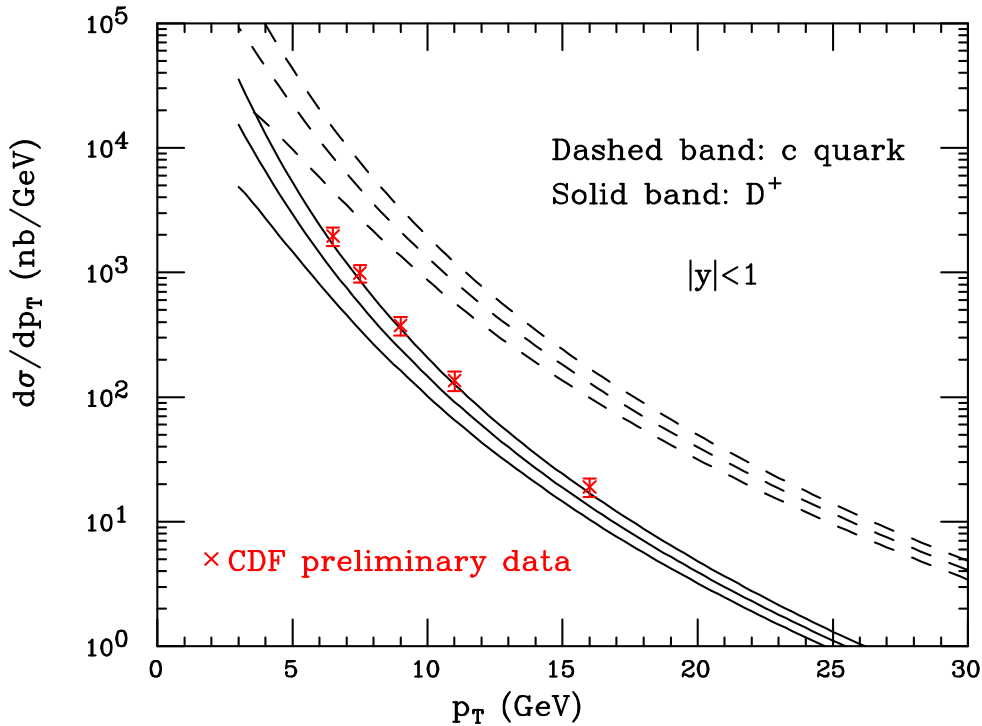}
\begin{center}
\caption{
Same as fig.~\protect\ref{fig:d0run2}, but for $D^{+}$ production.
}
\label{fig:dplusrun2}
\end{center}
\end{figure}

We show in
figs.~\ref{fig:d0run2},\ref{fig:dstarplusrun2},\ref{fig:dplusrun2} 
our predictions for the $d\sigma/d\pt~(|y|<1)$
cross sections for $D^0$, $D^{*+}$ and $D^+$ production at the Tevatron Run
II, obtained with the FONLL perturbative result combined with the 
non-perturbative fragmentation functions given in
eqs.~(\ref{eq:ffdstar}),(\ref{eq:ffdplus}) and (\ref{eq:ffdzero}) respectively.
The perturbative calculation is consistent with the one employed
in calculating $e^+e^-$ fragmentation when fitting the $D^*$ non-perturbative 
FF, and it is also shown for comparison.
The three curves in each plot are the lower, central and upper
values of an uncertainty band produced by varying independently the 
factorization and renormalization scales in the transverse mass 
region $\mt/2$ and
$2\mt$. The charm mass has been fixed here to $m = 1.5$~GeV, and the
CTEQ6M parton distribution functions set, corresponding to
$\Lambda^{(5)}$ = 0.226~GeV, has been used.
We have verified that accounting also for the effects of varying the QCD
scale in the range $0.192 < \Lambda^{(5)} < 0.241$~GeV 
and the charm mass in the range $1.2
< m < 1.8$~GeV does not produce a larger band than the one given by
factorization and renormalization scale variations\footnote{An
explanation of the limited sensitivity to the value of $m$  is that the
effects of mass variations are largely compensated by the need to also fit a
different non-perturbative parameter $r$ from the $e^+e^-$ ALEPH $D^*$
data, as $m$ also enters the evolution of the perturbative calculation. It
should also be stressed that the variation of either the mass or the
QCD scale is not computed 
fully consistently, because we are
not in a position to determine the effect of these variations on the
parton densities. We therefore content ourselves with computing them 
at fixed parton densities.}.

\begin{figure}[t]
\begin{center}
\epsfig{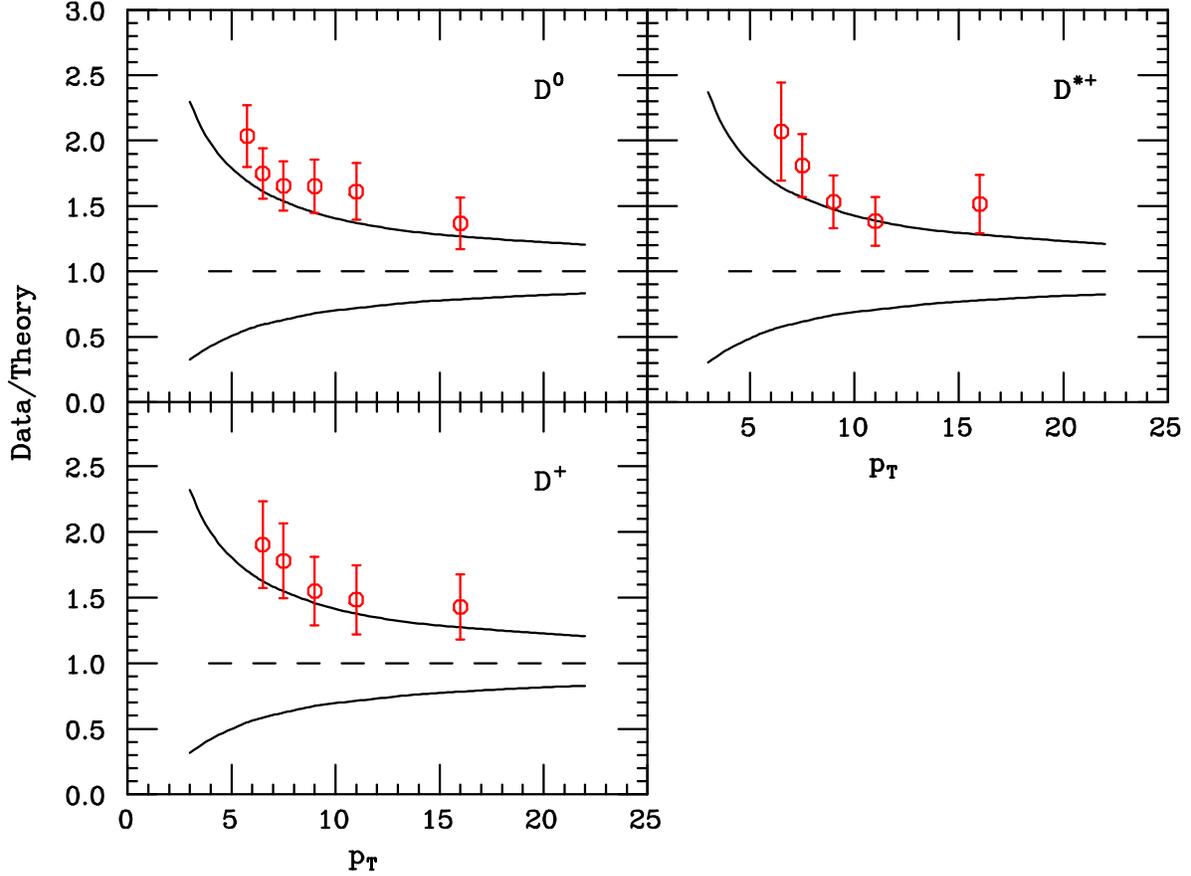}
\caption{The ratios between the experimental results and the central
theoretical predictions. Theoretical uncertainty bands are also shown,
defined as the ratio between the upper (lower) theoretical prediction
and the central value.}
\end{center}
\label{fig:ratios}
\end{figure}

The preliminary
data measured by the CDF Collaboration~\cite{Chen:2003qe} are also presented in the plots,
superimposed to the theoretical curves. They can be seen to be in fair
agreement with the predictions. The slopes in particular are very well
described, and they differ from that of the QCD prediction for charm
quarks, pointing to a correct determination of the softening effects
related to the non-perturbative fragmentation functions. Explicit numerical
values for the theoretical predictions are given in Table~\ref{table},
and the Data/Theory ratios are shown in fig.~\ref{fig:ratios} together 
with the theoretical
uncertainty bands, here defined as upper/lower theoretical prediction divided
by the central value.

\begin{table}[th]
\begin{center}
\begin{tabular}{|c|c|c|c|}
\hline
$p_T$ (GeV) & \multicolumn{3}{|c|}{$d\sigma/dp_T$ ($|y|<1$) (nb/GeV)} \\
\hline
            & $D^0$ & $D^{*+}$ & $D^+$ \\
\hline
5.75        & $.385^{+.266}_{-.171}$ $\times 10^4$&
              $.192^{+.140}_{-.089}$ $\times 10^4$&
	      $.170^{+.120}_{-.076}$ $\times 10^4$\\[2pt]
& 7837$\pm$220$\pm$884 & 
- & - \\
\hline
6.5         & $.232^{+.142}_{-.095}$ $\times 10^4$&
              $.117^{+.076}_{-.050}$ $\times 10^4$&
	      $.103^{+.065}_{-.043}$ $\times 10^4$\\[2pt]
&4056$\pm$93$\pm$441 &	      
2421$\pm$108$\pm$424  &
1961$\pm$69$\pm$332\\
\hline
7.5         & $.124^{+.067}_{-.046}$ $\times 10^4$&
              $.634^{+.359}_{-.245}$ $\times 10^3$&
	      $.554^{+.304}_{-.209}$ $\times 10^3$\\[2pt]
& 2052$\pm$58$\pm$227 &
1147$\pm$48$\pm$145&
986$\pm$28$\pm$156\\
\hline
9           & $.539^{+.242}_{-.173}$ $\times 10^3$&
              $.279^{+.132}_{-.093}$ $\times 10^3$&
	      $.242^{+.111}_{-.079}$ $\times 10^3$\\[2pt]
& 890$\pm$25$\pm$107&
427$\pm$16$\pm$54&
375$\pm$9$\pm$62\\
\hline
11          & $.203^{+.075}_{-.057}$ $\times 10^3$&
              $.107^{+.041}_{-.031}$ $\times 10^3$&
	      $.917^{+.346}_{-.263}$ $\times 10^2$\\[2pt]
& 327$\pm$15$\pm$41&
148$\pm$8$\pm$18&
136$\pm$4$\pm$24\\
\hline
16          & $.292^{+.078}_{-.062}$ $\times 10^2$&
              $.157^{+.044}_{-.035}$ $\times 10^2$&
	      $.133^{+.036}_{-.029}$ $\times 10^2$\\[2pt]
& 39.9$\pm$2.3$\pm$5.3&
23.8$\pm$1.3$\pm$3.2&
19.0$\pm$0.6$\pm$3.2\\
\hline
\end{tabular}
\caption{Theoretical predictions, with uncertainties, for the
differential cross sections of $D^0$, $D^{*+}$ and $D^+$ meson production
in $p\bar p$ collisions at the Tevatron Run II. The preliminary experimental
measurements from the CDF Collaboration~\protect\cite{Chen:2003qe} are also 
shown in the second lines at each $p_T$ value. The statistical and
systematic errors have been added in quadrature in the figures.}
\end{center}
\label{table}
\end{table}

The bands shown in the plots of 
figs.~\ref{fig:d0run2},\ref{fig:dstarplusrun2},\ref{fig:dplusrun2} do
not account for all the possible uncertainties. Besides the effects of
renormalization/factorization scales, mass and QCD scale variations,
previously discussed, one could also consider changing the Parton
Distribution Function set and/or modifying the model describing the
hadronization of the charm quark into the mesons. At the present level
of accuracy, however, we do not expect these issues to change significantly
the picture presented in the figures.

\section{Conclusions}\label{sec:conc}

We have presented predictions for $D^{*+}$, $D^+$ and $D^0$ transverse
momentum distributions as measured by the CDF Collaboration at the
Fermilab Tevatron Run II in $p\bar p$ collisions at a centre-of-mass
energy of 1960 GeV.

The calculation consists of next-to-leading log resummed QCD (matched to
a fixed next-to-leading order result) combined with non-perturbative
information describing the charm hadronization into the mesons, and
determined from moment-space analysis of $D^{*+}$ data measured by the
ALEPH Collaboration in $e^+e^-$ collisions.

The overall agreement is good, with the data lying on the upper limit
of the theoretical uncertainty band. This behaviour is similar to what
observed in $B$ mesons~\cite{Acosta:2001rz,Cacciari:2002pa} and
$b$-jets~\cite{Abbott:2000iv,Frixione:1997nh} production at the
Tevatron. This consistency is encouraging, and many improvements, like
tuning of scales and/or quark mass, adding higher order perturbative
contributions, or a  more accurate determination of non-perturbative
effects, could easily conspire to bring the theoretical prediction in
even better agreement with the experimental data.
It has been speculated that new physics
mechanisms may be needed in order to explain the measured excess
in the $b$ cross section \cite{Berger:2000mp}.
The presence of `new
physics' effects cannot certainly be ruled out at this stage.
However, the fact that both charm and bottom production data
bear similar relationship to the theoretical prediction
gives an indication that uncalculated QCD effects are a more
likely explanation of the discrepancy.

\vspace{1cm}
\noindent
{\bf Acknowledgments.} We wish to thank Chunhui Chen, Joseph Kroll, 
Michelangelo Mangano and Rolf Oldeman for useful conversations.

\bibliography{paper}
\end{document}